\documentstyle[12pt]{article}
\input epsf.sty
\renewcommand{\baselinestretch}{1.3}

 1

\catcode`\@=11 
\makeatletter
\def\@seccntformat#1{\csname the#1\endcsname.\hskip 1em}

\makeatother
\begin{document}
\thispagestyle{empty}
\begin{flushright}
{\footnotesize\renewcommand{\baselinestretch}{.75}
  SLAC--PUB--8160\\
June 1999\\
}
\end{flushright}

\vskip 0.3truecm
 
\begin{center}
{\bf \large A STUDY OF CORRELATIONS BETWEEN IDENTIFIED CHARGED HADRONS
IN HADRONIC $Z^0$ DECAYS$^*$}
 
\vspace {0.4cm}
 
 {\bf The SLD Collaboration$^{**}$}\\
Stanford Linear Accelerator Center \\
Stanford University, Stanford, CA~94309
 
\vspace{0.3cm}
 
\end{center}
 
\normalsize
 
 
\begin{center}
{\bf ABSTRACT }
\end{center}

{\small
\baselineskip=14pt
\noindent
We present a preliminary study of correlations in rapidity between pairs of
identified charged pions, kaons and protons using the entire SLD data sample
of 550,000 hadronic $Z^0$ decays.
Short range charge correlations are observed between all combinations of these
hadron species, indicating local conservation of quantum numbers and charge
ordering in the jet fragmentation process.
The rapidity range of this effect is found to be independent of particle
momentum.
A strong long-range $K^+$-$K^-$ correlation is observed at high-momentum 
and weaker long-range $\pi^+$-$\pi^-$, $\pi^+$-$K^-$ and p-$K^-$ and
p$\bar{\rm p}$
correlations are observed in light flavor events, providing new information on
leading particle production in $u$, $d$ and $s$ jets.
The long-range correlations observed in $c\bar{c}$ and $b\bar{b}$ events are
markedly different and consistent with expectations based on known decay
properties of the leading heavy hadrons.
In addition, the SLC electron beam polarization is used to tag the quark
hemisphere in each event, allowing the first study of rapidities signed such
that positive rapidity is along the quark rather than antiquark direction.
Distributions of ordered differences in signed rapidity 
between pairs of particles provide a direct probe of quantum number ordering
along the quark-antiquark axis and other new insights into the fragmentation
process.
}
 
\vskip 0.2truecm
 
\vfil
 
\noindent
Contributed to:  the
International Europhysics Conference on High Energy Physics,
15--21 July 1999, Tampere, Finland; Ref: 1\_187, 
and to the XIX$^{th}$ International Symposium on Lepton Photon Interactions,
August 9--14, 1999, Stanford, USA.

\vskip .3truecm

\noindent
$^*$Work supported by Department of Energy contract DE-AC03-76SF00515.

\eject

\renewcommand{\baselinestretch}{1.5}

\section{Introduction}

Correlations between particles produced in hadronic jets can be used to
probe details of the jet fragmentation process.
In $e^+e^-$ annihilations into hadrons, the total charge, strangeness, baryon
number, etc. of the final state particles in each event must be zero, and
it is interesting to ask whether the conservation of such quantum numbers
is local, or is longer-range in character.
For example, in the case of strangeness, one can ask whether a strange particle
and a corresponding antistrange particle tend to be produced ``close" to each
other within the event,
whether strange and antistrange particles are associated with the initial
$\bar{s}$ and $s$ quarks, respectively, in $s\bar{s}$ events
(or with the initial $u$ and $\bar{u}$ in $u\bar{u}$ events, etc.),
or whether strange and antistrange particles are distributed randomly
throughout the event.
Similar questions can be posed for other relevant quantum numbers.
Previous studies \cite{srcor} of differences in rapidity between associated
identified hadrons have shown that the conservation of charge, strangeness and
baryon number is predominantly local.
Most fragmentation models implicitly implement this feature, and
the form and range of such short-range correlations provide useful tests
of these models.
Short-range correlations can also arise from the decays of heavier hadrons, for
example the decay $\rho^0 \rightarrow \pi^+\pi^-$ will produce opposite-charge
pion pairs with a characteristic degree of locality.

Long-range correlations between particles of opposite charge and strangeness
in opposite jets of an event have also been observed \cite{lrcor}.
These can be understood in terms of leading particle production whereby the
higher-rapidity tracks in each jet tend to carry the quantum numbers of the
initial quark or antiquark.
In the case of $e^+e^- \rightarrow s\bar{s}$ events,
the $s$ and $\bar{s}$ quarks may hadronize, for example, into a high momentum 
$K^-$ and $K^+$, respectively, and there need be no other strange particles
in the event.
In $u\bar{u}$ and $d\bar{d}$ events, however, the locality of quantum number
conservation implies that a high-momentum strange-antistrange {\it pair} must
be produced in each jet, which will dilute any long-range correlation.
Nevertheless, improved measurements of long-range correlations may provide
better understanding of leading particle production.

The rapidity of a particle is typically defined with an arbitrary sign.
If a sign could be given to each measured rapidity such that, for example,
positive (negative) rapidity corresponds to the initial quark (antiquark)
direction, then one might probe more deeply into both leading and nonleading
particle production.
One could measure, for example, the extent to which a leading particle has
higher rapidity than its associated antiparticle,
and the extent to which low-momentum particles in jets remember the initial
quark/antiquark direction.

In this paper we present a study of correlations in rapidity between
identified charged pions, kaons and protons based on about 550,000 hadronic
$Z^0$ decays recorded by the SLD experiment at the SLAC Linear Collider.
Clean samples of identified particles were obtained using the Cherenkov Ring
Imaging Detector.
The hadronic event sample was divided into samples enriched in light-flavor
($e^+e^- \rightarrow u\bar{u}$, $d\bar{d}$ or $s\bar{s}$), $c\bar{c}$ and
$b\bar{b}$ events in order to study the effects of the decays of the leading
bottom and charmed hadrons.
In section 4 we present a study of short-range correlations between all
pair-combinations of
these hadron species as a function of hadron momentum.
We quantify the range of each correlation, and compare with the
predictions of the JETSET fragmentation model.
In section 5 we search for long-range correlations between these species.
In section 6 we exploit the SLC electron beam polarization to identify the quark
(vs. antiquark) hemisphere in each event and sign the rapidities such that
particles in the quark-tagged hemisphere have positive rapidity.
The signed rapidity distributions themselves provide new information on leading
particle production, and ordered rapidity differences between particle pairs
allow new and unique probes of the fragmentation process.

\section{Apparatus and Hadronic Event Selection}

A general description of the SLD can be found elsewhere~\cite{sld}.
The trigger and initial selection criteria for hadronic $Z^0$ decays are 
described in Ref.~\cite{sldalphas}.
This analysis used charged tracks measured in the Central Drift
Chamber (CDC)~\cite{cdc} and Vertex Detector (VXD)~\cite{vxd}, and identified
using the Cherenkov Ring Imaging Detector (CRID) \cite{crid}.
Momentum measurement is provided by a uniform axial magnetic field of 0.6T.
The CDC and VXD give a momentum resolution of
$\sigma_{p_{\perp}}/p_{\perp}$ = $0.01 \oplus 0.0026p_{\perp}$,
where $p_{\perp}$ is the track momentum transverse to the beam axis in
GeV/$c$.
One quarter of the data were taken with the original vertex detector (VXD2), and
the rest with the upgraded detector (VXD3).
In the plane normal to the beamline 
the centroid of the micron-sized SLC IP was reconstructed from tracks
in sets of approximately thirty sequential hadronic $Z^0$ decays to a precision 
of $\sigma_{IP}\simeq7$ $\mu$m for the VXD2 data and $\simeq$3 $\mu$m for the
VXD3 data.
Including the uncertainty on the IP position, the resolution on the 
charged track impact parameter ($\delta$) projected in the plane perpendicular
to the beamline is 
$\sigma_{\delta} =$11$\oplus$70/$(p \sin^{3/2}\theta)$ $\mu$m for VXD2 and
$\sigma_{\delta} =$8$\oplus$29/$(p \sin^{3/2}\theta)$ $\mu$m for VXD3, where
$\theta$ is the track polar angle with respect to the beamline. 
The CRID comprises two radiator systems that allow the identification of
charged pions
with high efficiency and purity in the momentum range 0.3--35 GeV/c, charged
kaons in the ranges 0.75--6 GeV/c and 9--35 GeV/c, and protons in the ranges
0.75--6 GeV/c and 10--46 GeV/c \cite{bfp}.
The event thrust axis~\cite{thrust} was calculated using energy clusters
measured in the Liquid Argon Calorimeter~\cite{lac}. 

A set of cuts was applied to the data to select well-measured tracks
and events well contained within the detector acceptance.
Charged tracks were required to have a distance of
closest approach transverse to the beam axis within 5 cm,
and within 10 cm along the axis from the measured IP,
as well as $|\cos \theta |< 0.80$, and $p_\perp > 0.15$ GeV/c.
Events were required to have a minimum of seven such tracks,
a thrust axis  polar angle w.r.t. the beamline, $\theta_T$,
within $|\cos\theta_T|<0.71$, and
a charged visible energy $E_{vis}$ of at least 20~GeV,
which was calculated from the selected tracks assigned the charged pion mass. 
The efficiency for selecting a well-contained $Z^0 \rightarrow q{\bar q}(g)$
event was estimated to be above 96\% independent of quark flavor.
The VXD, CDC and CRID were required to be operational, resulting in a 
selected sample of roughly 285,000 events, with an estimated
non-hadronic background contribution of $0.10 \pm 0.05\%$ dominated
by $Z^0 \rightarrow \tau^+\tau^-$ events.
 
Samples of events enriched in light and $b$ primary flavors were selected based
on charged track impact parameters $\delta$ with respect to the IP in the
plane transverse to the beam \cite{homer}.
For each event we define $n_{sig}$ as the number of tracks with
impact parameter greater than three times its estimated error,
$\delta > 3 \sigma_{\delta}$.
Events with $n_{sig}=0$
were assigned to the light-flavor sample and those with $n_{sig} \geq 4$
were assigned to the $b$-flavor sample; the remaining events were classified
as a $c$-flavor sample.
The light-, $c$- and $b$-tagged samples
comprised 166,000, 82,000 and 37,000 events,
respectively;
selection efficiencies and sample purities were estimated from our
Monte Carlo simulation and are listed in table \ref{tlveff}.

\begin{table}
\begin{center}
 \begin{tabular}{|l||c|c|c||c|c|c|} \hline
       & \multicolumn{3}{c||}{ } & \multicolumn{3}{c|}{ } \\ [-.5cm]
       & \multicolumn{3}{c||}{Efficiency for $Z^0 \rightarrow$}
       & \multicolumn{3}{c|}{Purity of $Z^0 \rightarrow$} \\
       & $u\bar{u},d\bar{d},s\bar{s}$ & $c\bar{c}$ & $b\bar{b}$ 
       & $u\bar{u},d\bar{d},s\bar{s}$ & $c\bar{c}$ & $b\bar{b}$  \\ [.1cm] \hline 
 &&&&&&\\[-.5cm] 
light-tag  & 0.846 & 0.338 & 0.034 & 0.881 & 0.106 & 0.013 \\
$c$-tag    & 0.153 & 0.617 & 0.401 & 0.320 & 0.388 & 0.292 \\
$b$-tag    & 0.001 & 0.045 & 0.565 & 0.005 & 0.064 & 0.931 \\[.1cm] \hline 
 \end{tabular}
\caption{\baselineskip=12pt \label{tlveff}
Tagging efficiencies for simulated events in the three flavor categories to be
tagged as light, $c$ or $b$.  The
three rightmost columns indicate the composition of each simulated tagged sample
assuming SM relative flavor production.}
\end{center}
\end{table}

Separate samples of hemispheres enriched in quark and antiquark jets
were selected by exploiting the large
electroweak forward-backward production asymmetry wrt the beam direction.
The event thrust axis was used to approximate the initial $q\bar{q}$ axis and
was signed such that its $z$-component was positive, $\hat{t}_z>0$.
Events in the central region of the detector, where the production asymmetry is
small, were removed by the requirement $|\hat{t}_z|>0.2$, leaving 235,000
events.
The quark-tagged hemisphere in events with left-(right-)handed electron beam
was defined to comprise the set of tracks with positive (negative) momentum
projection along the signed thrust axis.
The remaining tracks in each event
were defined to be in the antiquark-tagged hemisphere.
The sign and magnitude of the electron beam polarization were measured for every
event.
For the selected event sample, the average magnitude of the 
polarization was 0.73.
Using this value and assuming Standard Model couplings at 
tree-level, the purity of the quark-tagged sample is 0.73.

For the purpose of estimating the efficiency and purity of the event
flavor tagging and the particle identification,
we made use of a detailed Monte Carlo (MC) simulation of the detector.
The JETSET 7.4~\cite{jetset} event generator was used, with parameter
values tuned to hadronic $e^+e^-$ annihilation data~\cite{tune},
combined with a simulation of $B$-hadron decays
tuned~\cite{sldsim} to $\Upsilon(4S)$ data and a simulation of the SLD
based on GEANT 3.21~\cite{geant}.
Inclusive distributions of single-particle and event-topology observables
in hadronic events were found to be well described by the
simulation~\cite{sldalphas}.

\section{Identified Particle Selection}

The identification of charged tracks as pions, kaons or protons using the CRID
is described in detail in \cite{bfp}.
For this analysis we used a relatively loose identification algorithm, since the
presence of misidentified hadrons at the 10\% level has little effect on the
measured correlations.
Tracks with poor CRID information or that were likely to have scattered or
interacted before exiting the CRID were removed by requiring each track to have
at least 40 CDC hits,
at least one of which was at a radius of at least 92 cm, to extrapolate
through an active region of the appropriate CRID radiator and through a live
CRID TPC, and, in the case of the gas radiator,
to have fewer than four saturated
hits within the volume in which the gas ring is expected.  Approximately 85\% of
the tracks within the CRID acceptance satisfied these cuts.

Tracks identified in the calorimeters as electrons or muons \cite{emud} were
rejected, and for the remaining tracks log-likelihoods
\cite{bfp,llik} were calculated for each of the three charged hadron hypotheses
$i=\pi$, $K$ and p, and for each of the liquid and gas radiators.
A track was tagged as a hadron of species $i$ by the liquid (gas)
system if the log-likelihood for hypothesis $i$ exceeded both of the other
log-likelihoods by
at least 5 (3) units.  In addition, for those tracks with good information from
both the liquid and gas systems, the liquid and gas log-likelihoods were
added together, and a track was tagged by the combined system if the
log-likelihood for hypothesis $i$ exceeded both of the others by at least 3
units.
A track was identified as an $i$-hadron if it was tagged as type
$i$ by any of the liquid, gas or combined systems, and it was not tagged as
any other type by any other system.
The efficiencies for identifying accepted tracks are similar to those given in
\cite{bfp}: for charged pions there is roughly constant efficiency of about 80\%
in the
momentum range 0.5--25 GeV/c; charged kaons and protons have similar efficiency
except for a dip in the range 5--10 GeV/c.
The simulation was found to provide a good description of the momentum
distributions of the identified hadrons.

\begin{figure}
   \epsfxsize=4.5in
   \begin{center}\mbox{\epsffile{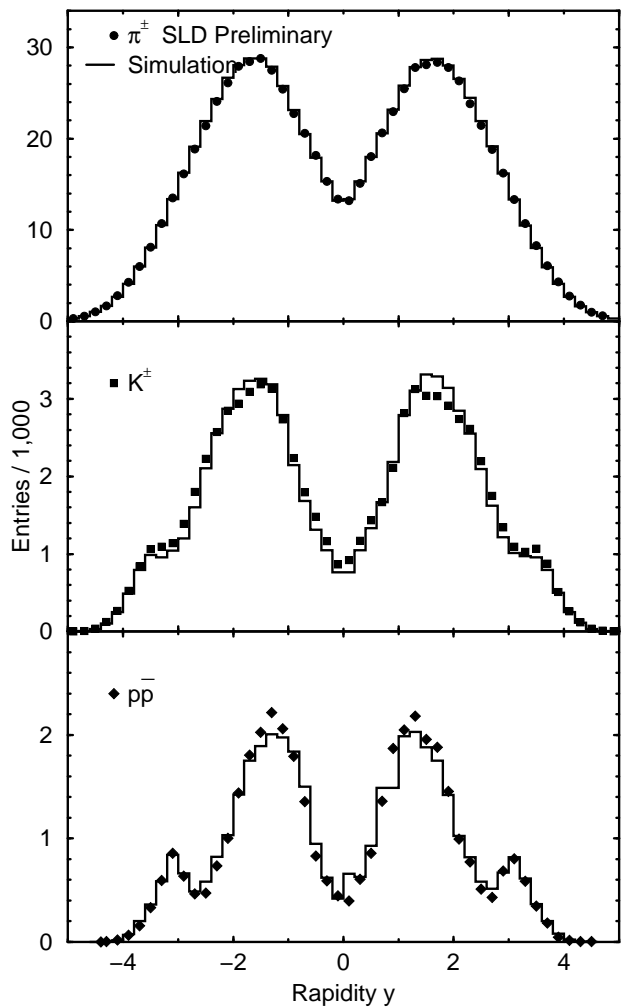}}\end{center}
  \caption{ 
 \baselineskip=14pt
 \label{frap}
Rapidity distributions for the identified pions, kaons and protons.
Also shown are the prediction of the Monte Carlo simulation.
}
\end{figure}

For each identified track the rapidity
$y=0.5\ln((E+p_\parallel)/(E-p_\parallel))$ was calculated using the measured
momentum and its projection $p_\parallel$ along the thrust axis,
and the appropriate hadron mass.
The distributions of rapidity are shown in fig. \ref{frap} for each
identified hadron species, along with prediction of the simulation.
Note that the overall sign of
the thrust axis vector, and therefore the sign of the rapidity, is arbitrary.
These distributions are not flat in the central region, but show structure due
to the momentum dependence of the CRID identification efficiency (see
\cite{bfp}).
The simulation provides a good qualitative description of the rapidity
distributions; the dependence of the identification
efficiency on momentum has only modest effects on the correlation studies,
as discussed below.

\begin{figure}
   \epsfxsize=6.6in
   \begin{center}\mbox{\epsffile{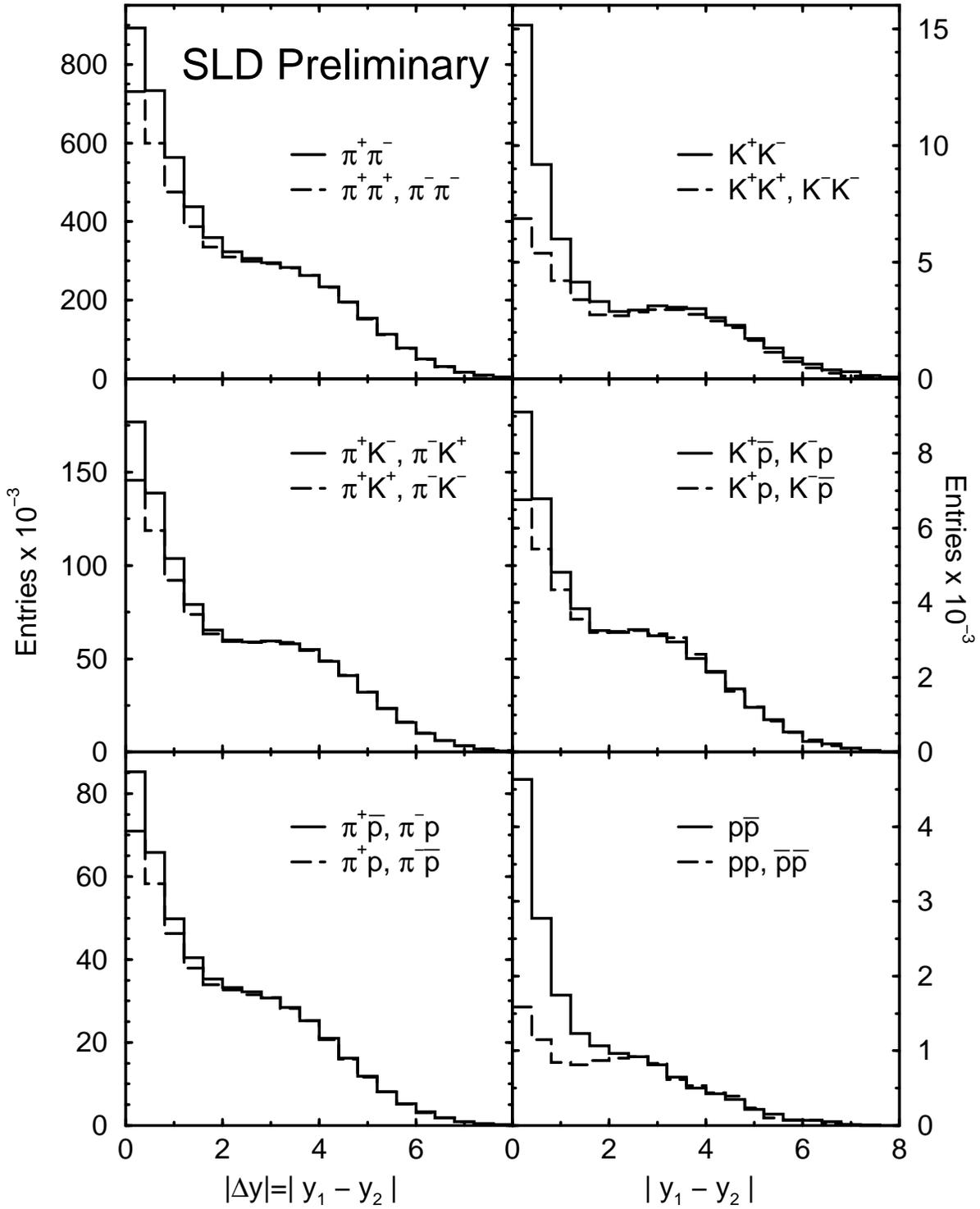}}\end{center}
  \caption{ 
 \baselineskip=14pt
 \label{fdrap}
Rapidity difference distributions for opposite-charge (histograms) and
same-charge (dashed histograms) pairs of identified pions, kaons and protons.
}
\end{figure}

The absolute value of the difference between the rapidities of each pair of
identified particles was calculated, and the distribution of this quantity is
shown in fig. \ref{fdrap} for each of the six possible pairs of hadron species.
In each case the distribution for those pairs with opposite charge is shown as
the solid histogram and that for pairs with the same charge is shown as the
dashed histogram.

\section{Short-Range Correlations}

For every type of hadron pair in fig. \ref{fdrap} there is an excess of
opposite-charge pairs over same-charge pairs at small values of the absolute
rapidity difference $|\Delta y|=|y_1-y_2|$.
We expect more opposite-charge than same-charge track pairs, as well as more
$K\bar{K}$ than $KK$/$\bar{K}\bar{K}$ pairs and more p$\bar{\rm p}$ than
pp/$\bar{\rm p}\bar{\rm p}$ pairs, due to 
conservation of electric charge, strangeness and baryon number, respectively.
In the case of $KK$ (pp) pairs the size of the excess is sensitive to the
relative fraction of strange-antistrange (baryon-antibaryon) pairs that are both
charged rather than neutral; if e.g. p-$\bar{\rm n}$ were the only type of
baryon-antibaryon pair produced, there would be no such excess.
The fact that the excess of opposite-charge $KK$ and pp pairs peaks at low
values of $|\Delta y|$ indicates that conservation of strangeness and baryon
number, respectively, is local, as has been observed previously \cite{srcor}.

\begin{figure}
   \epsfxsize=6.2in
   \begin{center}\mbox{\epsffile{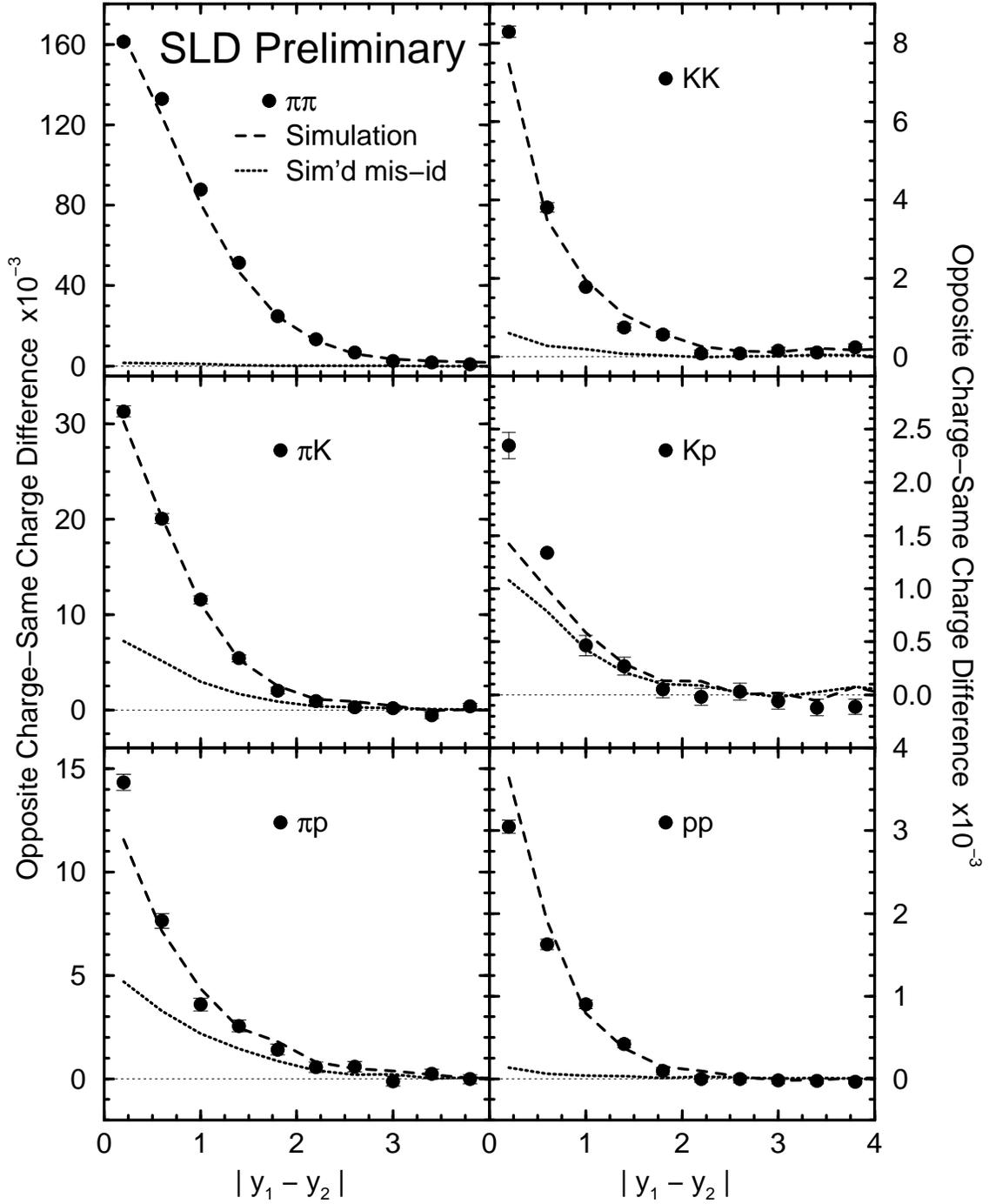}}\end{center}
  \caption{ 
 \baselineskip=14pt
 \label{fddraps}
Differences between the $|\Delta y|$ distributions for opposite-charge and
same-charge pairs in the small $|\Delta y|$ region.
The errors are statistical only.
Also shown are the predictions of the Monte Carlo simulation
for all detected pairs (dashed lines) and those pairs in which one or both
particles were misidentified (dotted lines).
There are normalization uncertainties of $\sim$5\% on the total predictions and
$\sim$20\% on the component due to misidentified particles.
}
\end{figure}

In order to study these short-range correlations in more detail, we assumed that
the tracks in each same-charge pair are unassociated, and subtracted their
$|\Delta y|$
distributions from those of the respective opposite-charge pairs.
These differences are shown for the low $|\Delta y|$ region
in fig. \ref{fddraps}, and are seen to differ significantly from each other
in form.
Monte Carlo studies indicate that these differences are not due to acceptance,
momentum dependence of the particle identification efficiency or to background
from misidentified particles.
Also shown in fig. \ref{fddraps} are the predictions of the simulation for this
difference.
In all cases, the simulation gives a reasonable description of the form of the
difference, although there are small differences in the details of the form
for $\pi\pi$, $\pi$p, $KK$ and $\pi$p pairs.
The predictions for the amplitudes have $\sim$5\% normalization uncertainties
(not shown) due to uncertainties in the particle identification efficiencies.
The predicted amplitudes are thus consistent with the data, except that for the
$K$p correlation, which is low by about 40\%.

Pairs of tracks from the decays of resonances
(e.g. $\rho^0 \rightarrow \pi^+\pi^-$) contribute to varying degrees for
all pair combinations;  each decay mode gives a $|\Delta y|$ difference
distribution with a
characteristic form, and collectively they have a substantial influence on
the overall form of the distributions.
Since the resonance production rates are adjustable in the simulation, we
suspect that it is able to provide an adequate description of the data.

Pairs in which one or both particles are misidentified are an important source
of background, and lead to a convergence in the forms of the distributions for
different pair combinations.
The simulated differences for those pairs in which at least one track is
misidentified are also shown in fig. \ref{fddraps}, and are seen to contribute
a sizeable fraction of the observed differences in the cases of unlike particles
($\pi K$, $\pi$p and $K$p).
There are normalization uncertainties on these predictions of $\sim$20\% due to
uncertainties in the misidentification rates.
Excellent particle identification is required to demonstrate, as seen in the
figure, that there are excesses at short range for true pairs of these three
combinations,
indicating local conservation of electric charge between these different
particle species and suggesting that there is
charge ordering among hadrons of all species in the fragmentation process.
The prediction of the simulation for the observed $K$p correlation is dominated
by misidentification;  the prediction for the true correlation is therefore low
by a factor of $\sim$3.

We have studied these short-range correlations in 6 bins of the momentum of
the heavier (higher momentum) particle in the case of $\pi K$, $\pi$p and
$K$p ($\pi\pi$, $KK$ and pp) pairs.
We find the quality of the simulated description of the data to be
independent of the momentum bin.
In order to quantify the range of each correlation, we fitted an ad hoc
function, the sum of two Gaussians, to each difference in each bin
over the range $0<|\Delta y|<3$ units.
The centers of both Gaussians were fixed to zero, and the amplitude (width) of
the wider Gaussian was fixed to 0.4 (2.2) times that of the narrower Gaussian,
leaving the amplitude and width of the narrower Gaussian as free parameters.
We used the width as a measure of the range of the correlation.
This function provided a reasonable qualitative description of both the data and
simulation for all pair combinations in all momentum bins.

\begin{figure}
   \epsfxsize=6.6in
   \begin{center}\mbox{\epsffile{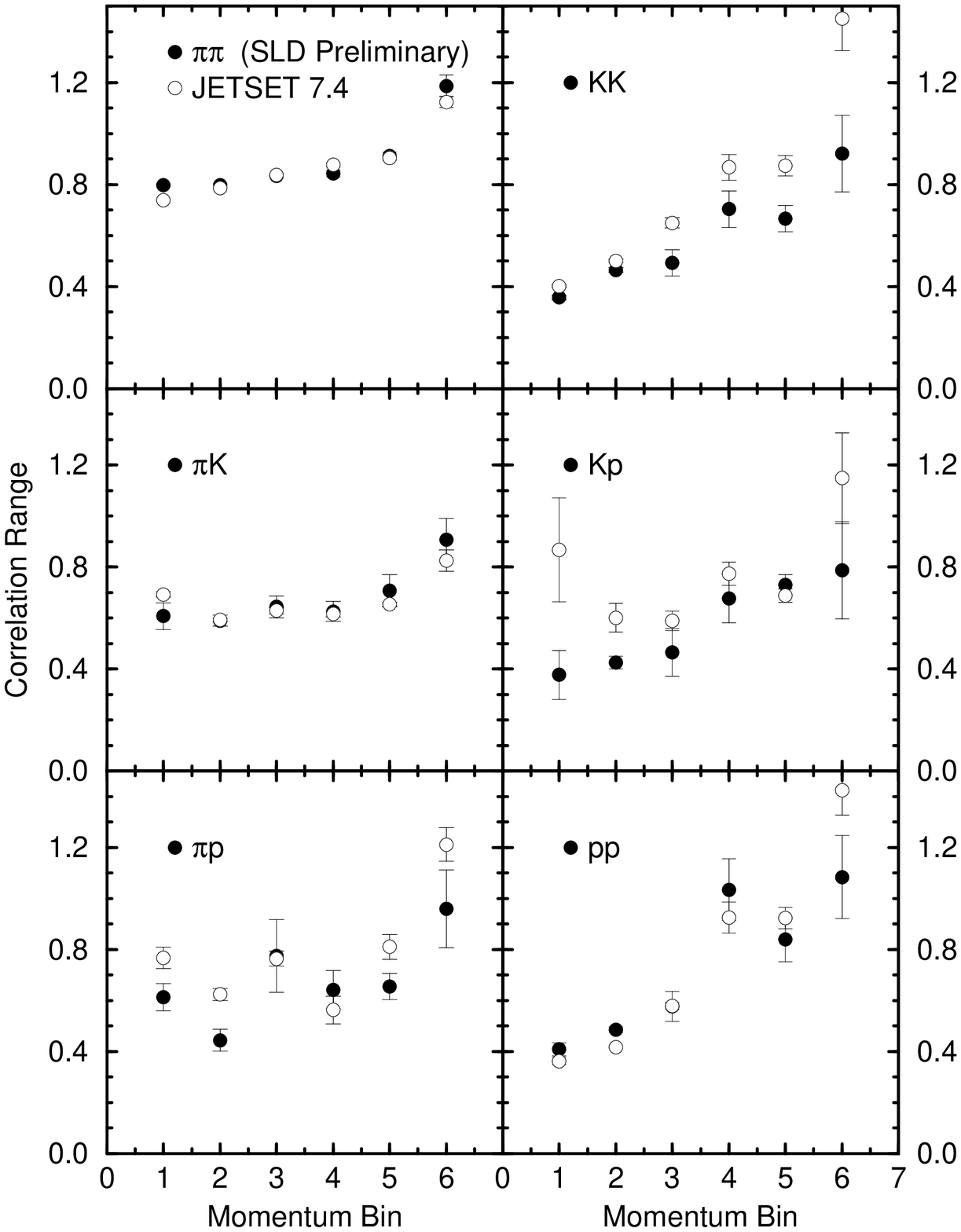}}\end{center}
  \caption{ 
 \baselineskip=14pt
 \label{fddwid}
Range (see text) of the observed short-range correlations in the data (dots) and
simulation (open circles), in six bins of momentum for each pair combination.
}
\end{figure} 

The fitted widths are shown in fig. \ref{fddwid} as a function of momentum for
each pair combination.
If the fragmentation process is scale invariant, then we expect the range to be
independent of momentum, except for biases introduced by the bin edges and the
particle identification.
In the case of $\pi\pi$, the range is constant within ten percent except for
the highest momentum bin.
Significant momentum dependence is observed for $\pi$p, $KK$, $K$p and pp
pairs, however this
dependence is reproduced by the simulation,
with the possible exception of differences at low momentum for $\pi$p and $K$p
pairs, and at high momentum for $KK$ pairs.
In the simulation, there is a $\sim$10\% momentum dependence due to binning and
particle decays.
The large slopes and point-to-point structure are due
to the momentum dependence of the particle identification efficiencies.
Thus the hypothesis of momentum-independence of the range of the short-range
rapidity correlation for a given pair combination is consistent with the data,
within the context of the JETSET model.

The qualitative results presented above were found to be independent of the
primary flavor of the event, as were the quality of the predictions of the
simulation and the feature of scale independence.
This is expected, as the only difference at short range should be due to the
decay properties of the leading $B$ and $D$ hadrons;  we do observe some small
differences between the flavor-tagged samples in the measured ranges of the
correlations, which are reproduced by the simulation.

\section{Long-Range Correlations}

\begin{figure}
   \epsfxsize=6.6in
   \begin{center}\mbox{\epsffile{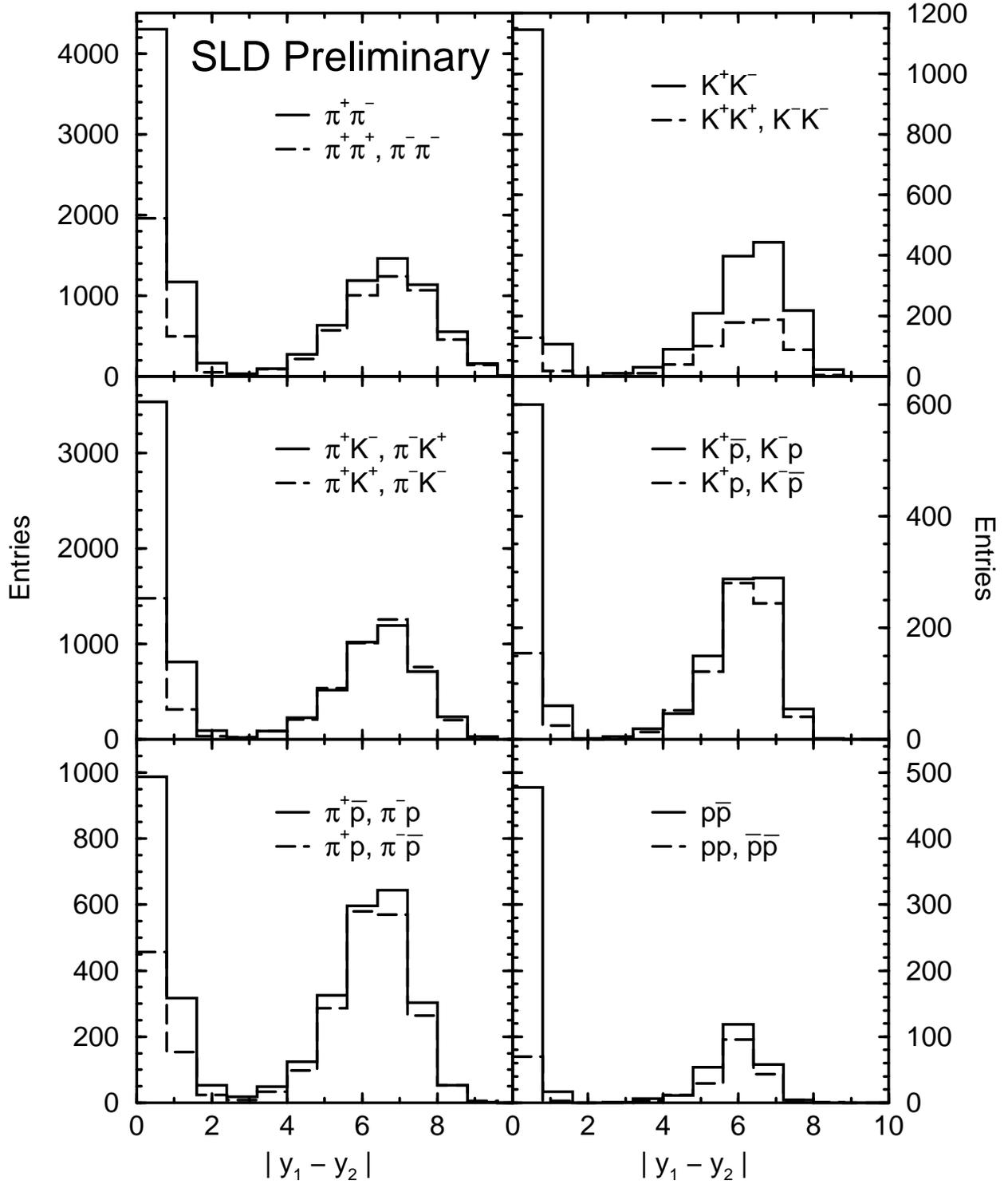}}\end{center}
  \caption{ 
 \baselineskip=14pt
 \label{fdrap9}
Observed $|\Delta y|$ distributions for opposite-charge (histograms) and
same-charge (dashed histograms) pairs in which both tracks have $p>9$ GeV/c.
}
\end{figure}

We next searched for long-range correlations between all pair combinations.
In fig. \ref{fdrap}, a difference between opposite-charge and same-charge pairs
at high $|\Delta y|$ is visible only in the case of $KK$, and even here the
background from uncorrelated pairs is dominant.
Since we expect long-range correlations from leading particle production to be
relatively more important at high momentum, we have studied these differences
for pairs in which both tracks have momentum $p>9$ GeV/c.
The corresponding $|\Delta y|$ distributions for each of the six pair
combinations are
shown in fig. \ref{fdrap9}.
For these high-momentum pairs, there is a separation between pairs in the
same jet ($|\Delta y|<2.5$) and those in opposite jets ($|\Delta y|>4$).
At short range there is a large excess of opposite-charge pairs over same-charge
pairs for all pair combinations, confirming that locality holds even at the highest
momenta.
At long range, there are clear correlations for $\pi\pi$ and $KK$ pairs, as well
as significant correlations for $\pi$p, $K$p and pp pairs.

\begin{figure}
   \epsfxsize=6.6in
   \begin{center}\mbox{\epsffile{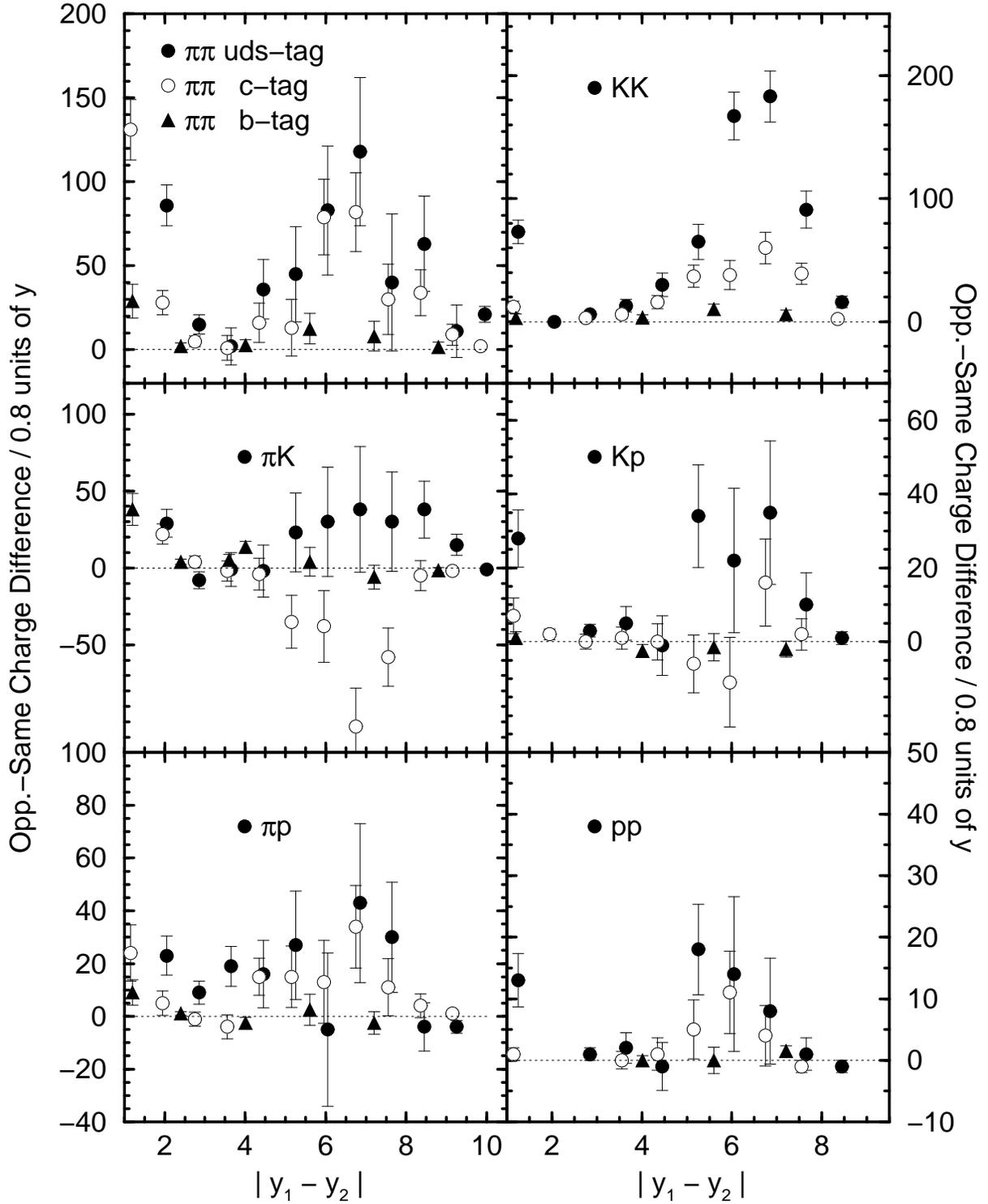}}\end{center}
  \caption{ 
 \baselineskip=14pt
 \label{fddr9fl}
Observed differences between the $|\Delta y|$ distributions for opposite-charge
and same-charge pairs in which both tracks have $p>9$ GeV/c, for
the light-(dots), $c$- (open circles) and $b$- (triangles) tagged samples.
}
\end{figure}

These long-range correlations are found to be strongly flavor-dependent.
Figure \ref{fddr9fl} shows the differences between the opposite-charge and
same-charge pairs in the high $|\Delta y|$ region for each of the flavor-tagged 
samples.  The $b\bar{b}$ events are seen to contribute very little to the
difference for any pair combination,
primarily because there are relatively few tracks with
such high momentum in these events.  Light flavor and $c\bar{c}$ events
contribute roughly equally to the observed correlations for $\pi\pi$, $\pi$p
and pp pairs;  light-flavor events dominate the $K$p correlation.
In the case of $\pi$K pairs, there is a strong anticorrelation in
$c\bar{c}$ events along with a correlation in light-flavor events, both of which
were invisible in the flavor-inclusive sample.

The integrated differences for $|\Delta y|>4$ are given for each flavor-tagged
sample in table \ref{tlrcor}.
The predictions of the simulation are also given and are generally consistent
with the data.
The simulation also describes the rapidity dependence in fig. \ref{fdrap9}
within the experimental uncertainties.

\begin{table}
\begin{center}
 \begin{tabular}{|c||r|r|r||r|r|r|} \hline
       & \multicolumn{3}{c||}{ } & \multicolumn{3}{c|}{ } \\ [-.5cm]
 Pair  & \multicolumn{3}{c||}{Data} & \multicolumn{3}{c|}{Simulation} \\
 Type  & $uds$-tag & $c$-tag & $b$-tag & $uds$-tag & $c$-tag & $b$-tag \\ [.1cm] \hline 
 &&&&&&\\[-.5cm] 
 $\pi\pi$ & 398$\pm$85 &   265$\pm$46 &   50$\pm$27
          & 488$\pm$42 &   241$\pm$23 &   41$\pm$10 \\
 $\pi K$  & 171$\pm$73 &$-$245$\pm$45 &   15$\pm$25
          &  43$\pm$34 &$-$169$\pm$20 &   20$\pm$ 9 \\
 $\pi$p   & 103$\pm$53 &    93$\pm$28 & $-$5$\pm$15
          & 145$\pm$25 &    44$\pm$13 &    8$\pm$ 5 \\
 $KK$     & 552$\pm$33 &   192$\pm$20 &   37$\pm$12
          & 449$\pm$16 &   190$\pm$10 &   28$\pm$ 4 \\
 $K$p     & 101$\pm$33 &     1$\pm$20 &$-$11$\pm$10
          &  76$\pm$15 &     1$\pm$ 9 & $-$6$\pm$ 3 \\
 pp       &  39$\pm$17 &    20$\pm$10 &    2$\pm$10
          &  18$\pm$ 7 &     6$\pm$ 4 &    1$\pm$ 1 \\ \hline
 \end{tabular}
\caption{
\baselineskip=12pt
\label{tlrcor}
Observed differences between the numbers of opposite-charge and same-charge
pairs in which both tracks had $p>9$ GeV/c and $|\Delta y|>4$, in events in each
of the three flavor-tagged samples.
Also shown are the predictions of the Monte Carlo simulation.}
\end{center}
\end{table}

\section{Signed Rapidities and Correlations}

We next tagged the quark (vs. antiquark)
direction in each hadronic event using the
electron beam polarization for that event, exploiting the large forward-backward
quark production asymmetry in $Z^0$ decays.  If the beam was left-(right-)handed
then the thrust axis was signed such that $\cos\theta_T$ was positive
(negative).
Events with $|\cos\theta_T|<0.15$ were removed, as the production asymmetry is
small in this region.
The probability to tag the quark direction correctly in these events was 73\%.

\begin{figure}
   \epsfxsize=5.0in
   \begin{center}\mbox{\epsffile{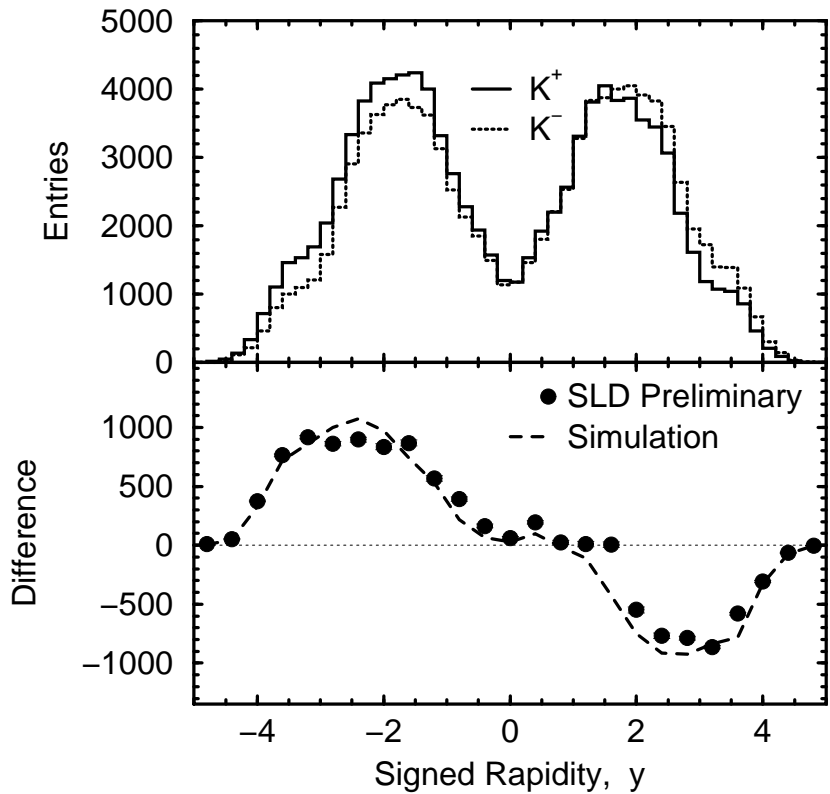}}\end{center}
  \caption{ 
 \baselineskip=14pt
 \label{fsrapk}
Distributions (top) of the rapidity with respect to the signed thrust axis for
positively (histogram) and negatively (dashed histograms) charged kaons.
The difference (bottom) between these two distributions compared with
the prediction of the Monte Carlo simulation.
}
\end{figure}

The rapidity of a particle with respect to the signed thrust axis is naturally
signed such that positive rapidity corresponds to the hemisphere in the
tagged direction of the initial quark,
and negative rapidity corresponds to the tagged antiquark hemisphere.
The signed rapidity distributions for identified $K^+$ and $K^-$ are shown in
fig. \ref{fsrapk}.  There is a clear difference between the two, with more $K^-$
than $K^+$ in the quark hemisphere, as expected due to leading $K^-$ produced
in $s$ quark jets \cite{bfp,lpprl}.
The difference between the two distributions is also shown in the figure and is
compared with the prediction of the simulation, which is consistent with the
data.

\begin{figure}
   \epsfxsize=6.6in
   \begin{center}\mbox{\epsffile{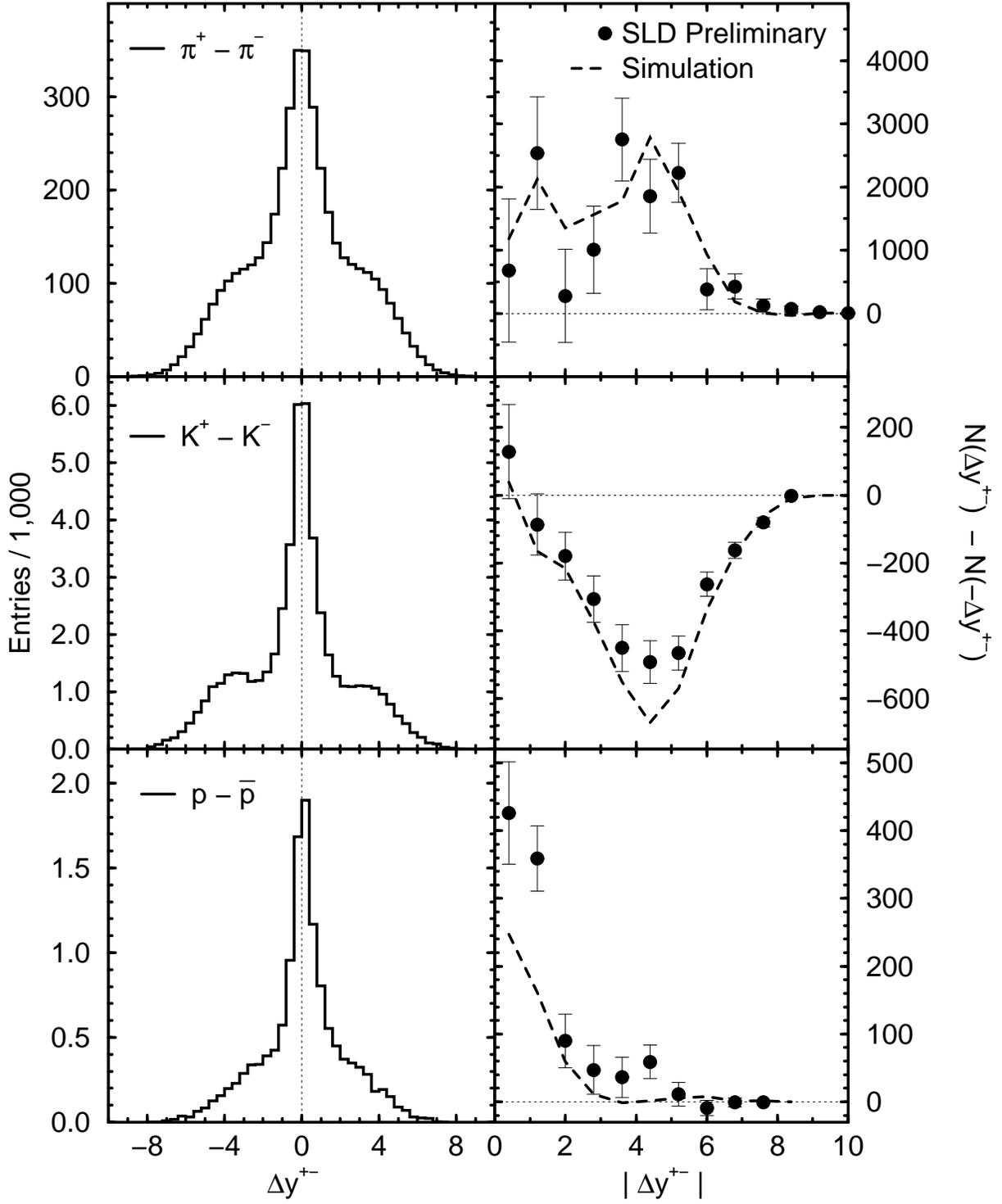}}\end{center}
  \caption{ 
 \baselineskip=14pt
 \label{fdsrapl}
Distributions (left) of the difference between the signed rapidities of
positively and negatively charged identified hadrons of the same type.
The differences (right) between the right and left sides of the distributions,
compared with the prediction of the Monte Carlo simulation.
}
\end{figure}

For pairs of identified particles, one can define an ordered rapidity
difference.
For particle-antiparticle pairs, we define $\Delta y^{+-} = y_+ - y_-$ as the
difference between the signed rapidities of the positively charged particle
and the negatively charged particle.
In fig. \ref{fdsrapl} we show the distribution of $\Delta y^{+-}$ for
$\pi^+\pi^-$, $K^+K^-$ and p$\bar{\rm p}$ pairs.
Asymmetries in these distributions are indications of ordering along the event
axis, and the differences between the positive and negative sides of these
distributions are also shown.
The negative difference at high $|\Delta y^{+-}|$ for the $K^+K^-$ pairs can be
attributed to
the fact that leading kaons are produced predominantly in $s\bar{s}$ events.
Similar but smaller effects for expected for $\pi^+\pi^-$ and
p$\bar{\rm p}$ pairs, and the positive difference in the latter at
$|\Delta y| \approx 4$ may be attributed to this effect.
For $\pi^+\pi^-$ pairs we observe a large positive difference at high
$|\Delta y^{+-}|$ rather than the expected small negative difference, which is
due entirely to $c\bar{c}$ events (see below).
The predictions of the simulation are also shown and are consistent
with the data at high $|\Delta y^{+-}|$.

\begin{figure}
   \epsfxsize=6.0in
   \begin{center}\mbox{\epsffile{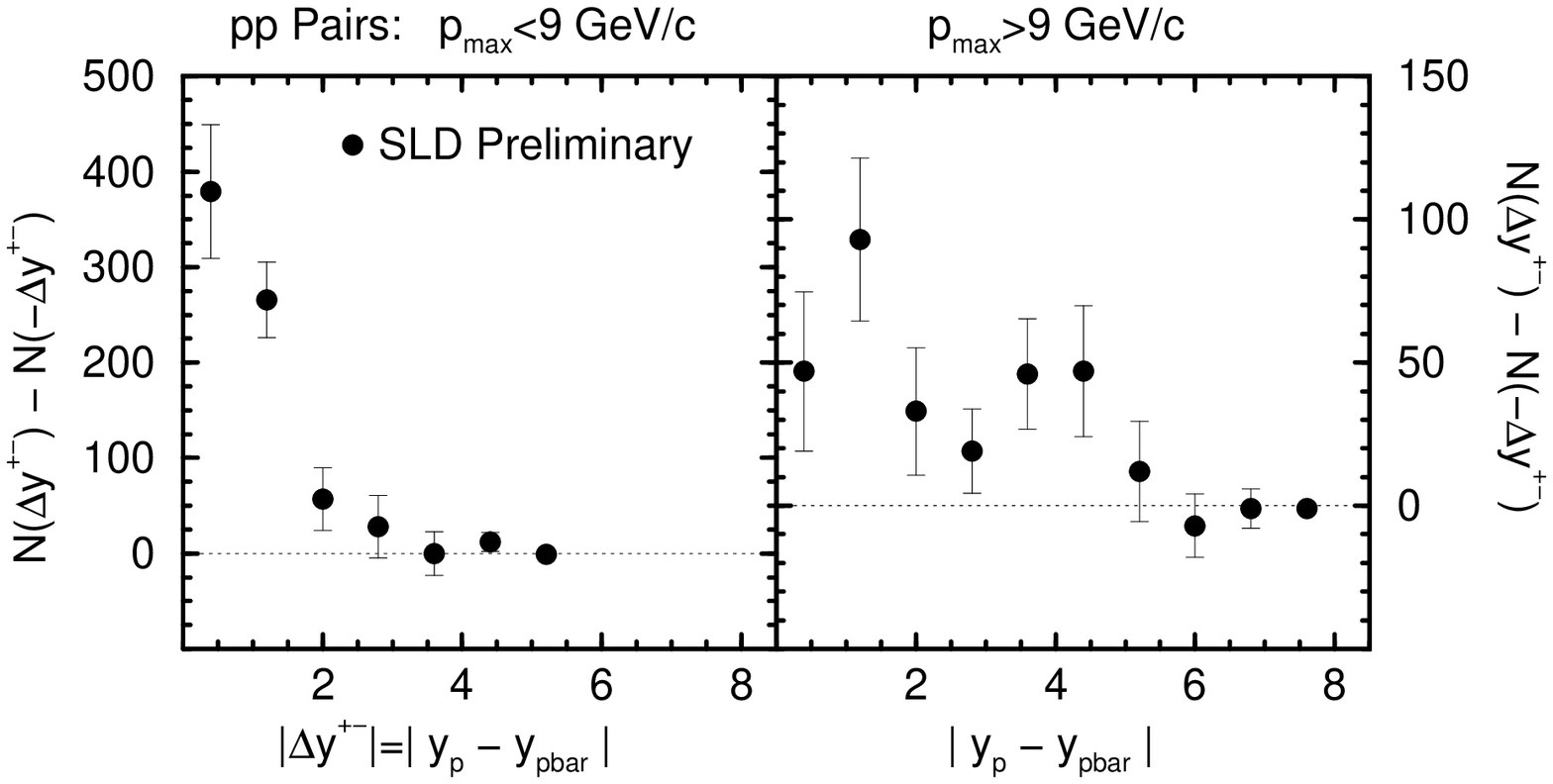}}\end{center}
  \caption{ 
 \baselineskip=14pt
 \label{fdsrpp}
Differences between the positive and negative sides of the distributions
of $y_{\rm p}-y_{\bar{\rm p}}$ at low (left) and high (right) momentum of the
higher-momentum track.
}
\end{figure}

The positive difference in the lowest $|\Delta y^{+-}|$ bins for the
p$\bar{\rm p}$ pairs indicates that the baryon in an associated
baryon-antibaryon pair follows the quark direction more closely than the
antibaryon.
This could be due to leading baryon production and/or to baryon number ordering
along the entire fragmentation chain.
We find a significant effect in all six of our momentum bins, and that the bulk
of the observed difference occurs at low momentum (see fig. \ref{fdsrpp}).
We therefore conclude that both of these effects contribute;
this is the first direct observation of baryon number ordering along the
entire chain.
Figure \ref{fdsrpp} also shows that the effect at long range noted above is
confined to high-momentum pairs, as expected if due to leading baryon
production.
The prediction of the simulation is low by a factor of two at low
$|\Delta y^{+-}|$ in both fig. \ref{fdsrapl} and in the low $p_{max}$ side of
fig. \ref{fdsrpp};  it is consistent with the high $p_{max}$ data.

\begin{figure}
   \epsfxsize=6.0in
   \begin{center}\mbox{\epsffile{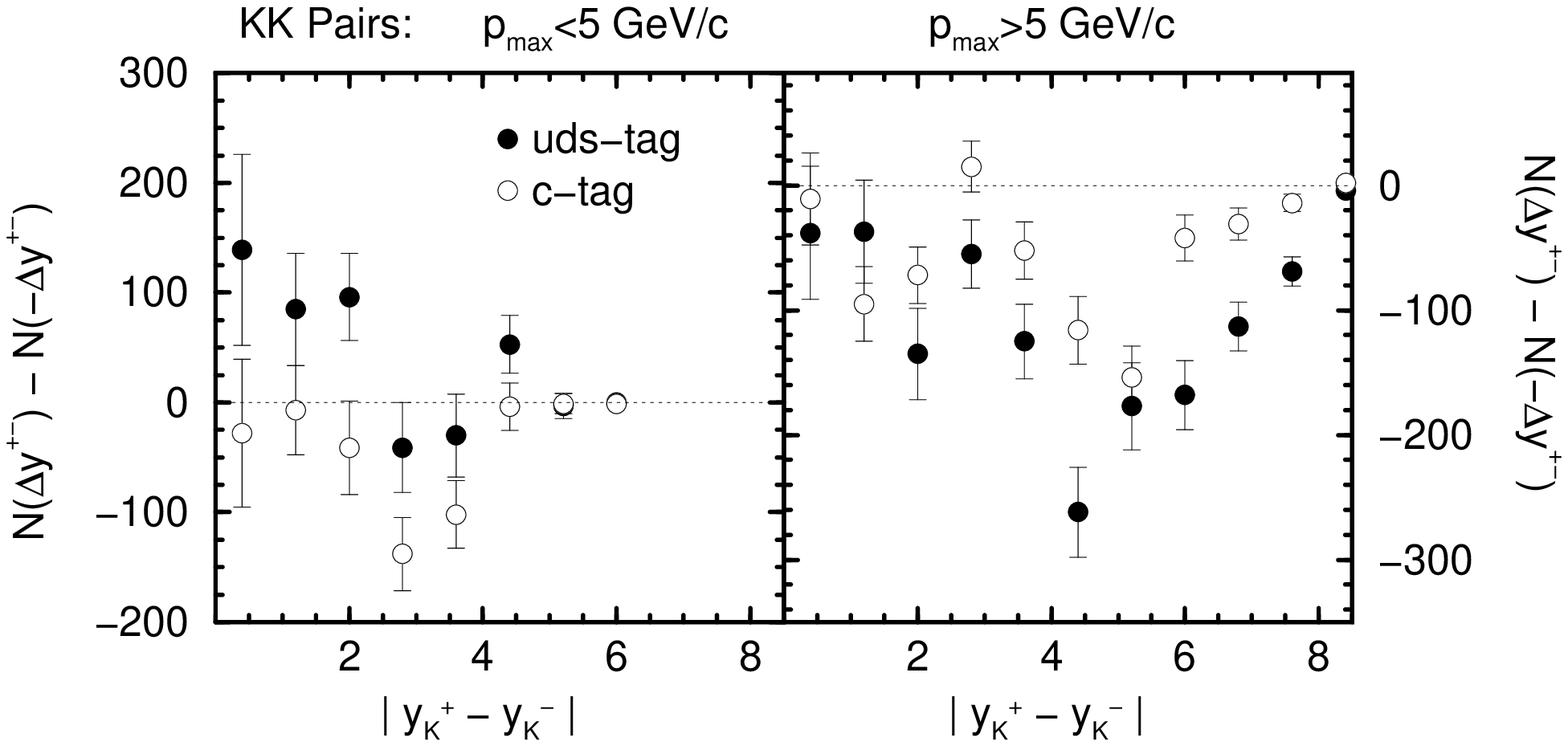}}\end{center}
  \caption{ 
 \baselineskip=14pt
 \label{fdsrkk}
Differences between the positive and negative sides of the distribution
of $y_{K^+}-y_{K^-}$ for pairs
produced at low (left) and high (right) momentum,
in the light- (dots) and $c$-tagged (open circles) samples.
}
\end{figure}

In fig. \ref{fdsrkk} we show the difference for $K^+K^-$ pairs in two bins of
$p_{max}$ for the light- and $c$-tagged samples separately.  For large
$p_{max}$ one can see differences in amplitude and in $|\Delta y|$ between the
contributions from these two flavor-tagged samples.
At low $p_{max}$ there is a positive difference of three standard deviations
at low $|\Delta y|$ for the light-flavor sample,
a possible indication of strangeness ordering along
the quark-antiquark axis for associated $K^+K^-$ pairs, similar to the baryon
number ordering observed for p$\bar{\rm p}$ pairs.
Such an effect is expected to be diluted at high momentum by an effect of the
opposite sign due to triplets of high-momentum kaons produced in $s\bar{s}$
events.
We do not observe such a signal in the heavy flavor events, possibly due to
limited statistics, dilution due to heavy hadron decays that include a $K^+$
and a $K^-$,
and/or dilution due to the negative difference from opposite hemisphere
pairs, that populate a lower $|\Delta y|$ region in heavy-flavor events than in
light-flavor events.

\begin{figure}
   \epsfxsize=6.0in
   \begin{center}\mbox{\epsffile{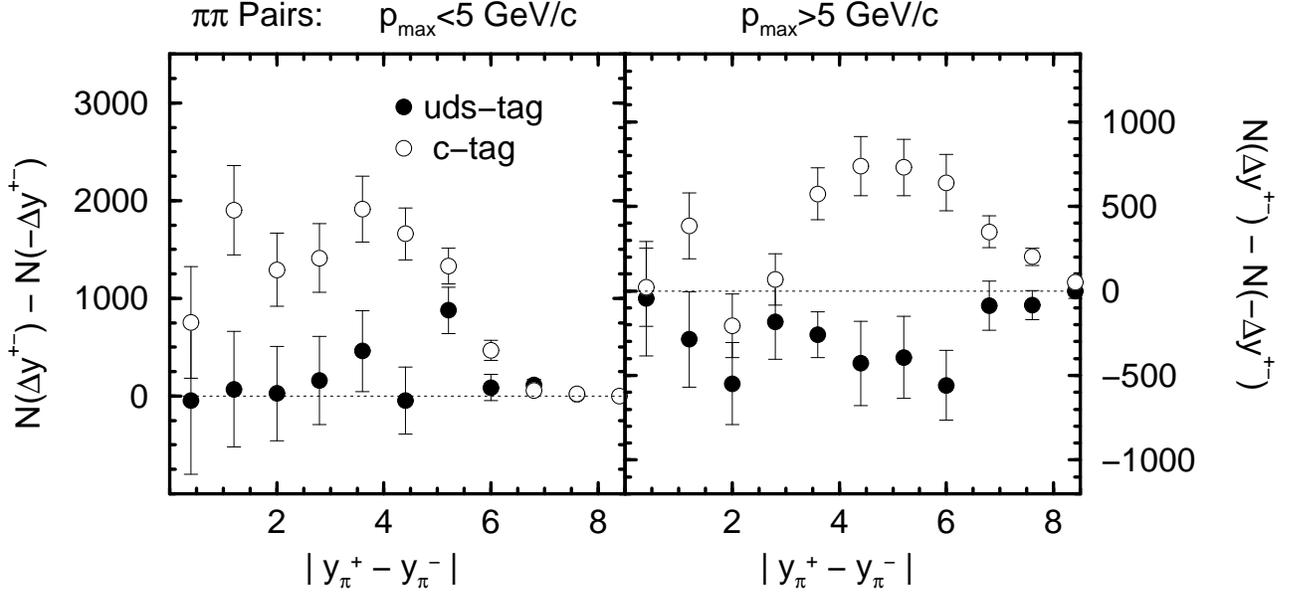}}\end{center}
  \caption{ 
 \baselineskip=14pt
 \label{fdsrpi}
Differences between the positive and negative sides of the distribution
of $y_{\pi^+}-y_{\pi^-}$ at low (left) and high (right) momentum,
in the light- (dots) and $c$-tagged (open circles) samples.
}
\end{figure}

In fig. \ref{fdsrpi} we show the difference for $\pi^+\pi^-$ pairs in two bins
of $p_{max}$ for the light-flavor and $c\bar{c}$ samples separately.
For large $p_{max}$ we observe a negative difference at high $|\Delta y|$
in the light-flavor sample, as expected from leading pion production.
In the $c\bar{c}$ events, there are positive differences in both bins of
$p_{max}$ at high $|\Delta y^{+-}|$ and also in the low $p_{max}$ bin at low
$|\Delta y^{+-}|$, all of 
which can be attributed to pairs involving a $\pi^+$ from the $D$
meson decay and a $\pi^-$ from the $\bar{D}$ meson decay.
The predictions of the simulation are consistent with these data.

\begin{figure}
   \epsfxsize=6.6in
   \begin{center}\mbox{\epsffile{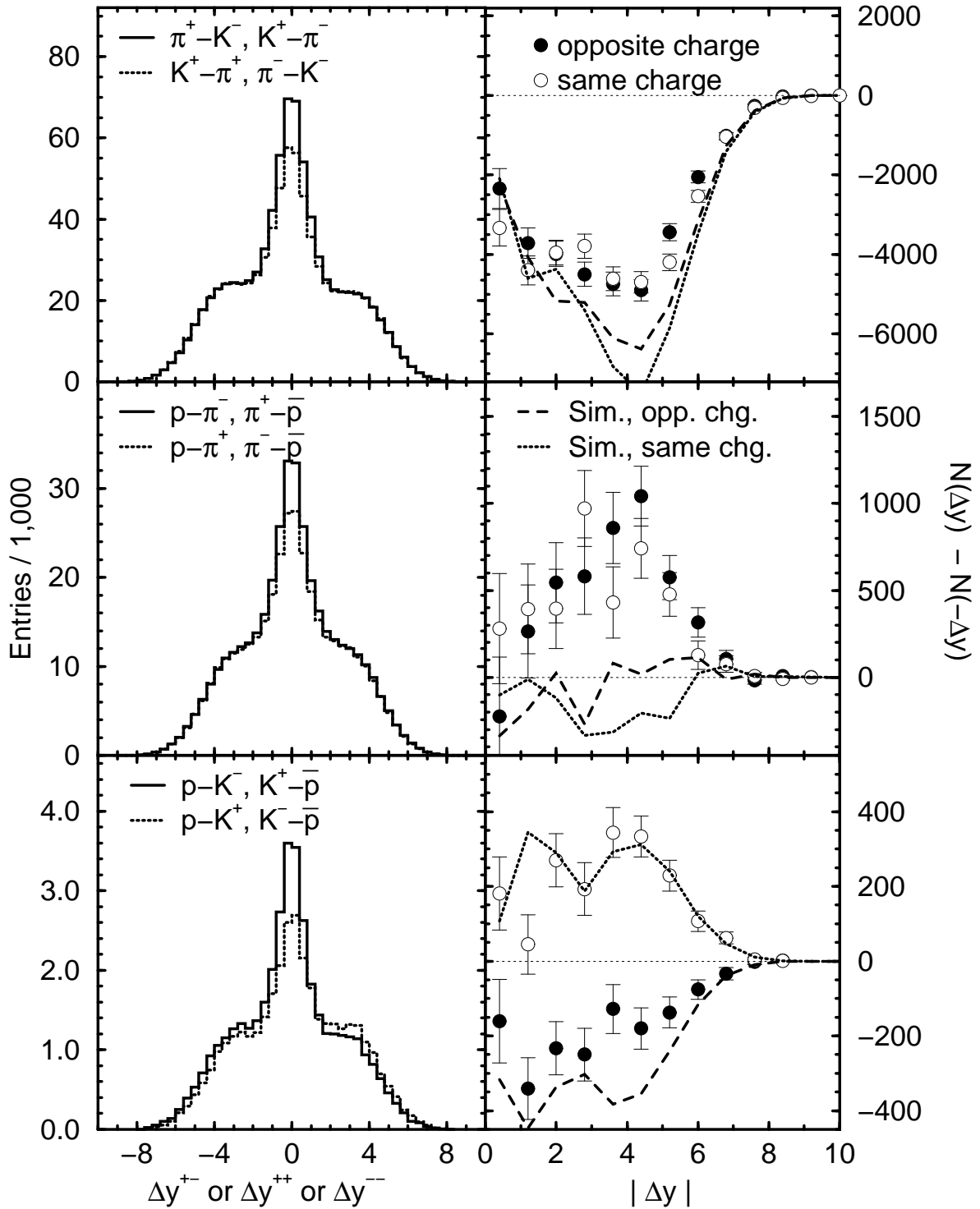}}\end{center}
  \caption{ 
 \baselineskip=14pt
 \label{fdsrapu}
Distributions (left) of the difference between the signed rapidities of
positively and negatively charged identified hadrons of different types.
The differences (right) between the right and left sides of the distributions,
compared with the predictions of the Monte Carlo simulation.
}
\end{figure}

For unlike particles, both opposite-charge and same-charge pairs may be of
interest.
We define the ordered difference as the rapidity of the heavier particle minus
that of the lighter particle multiplied by the charge of the heavier particle.
That is, $\Delta y^{+-} = y_{K^+} - y_{\pi^-}$ or $y_{\pi^-} - y_{K^+}$, 
$\Delta y^{++} = y_{K^+} - y_{\pi^+}$,
$\Delta y^{--} = y_{\pi^-} - y_{K^-}$, etc.
In fig. \ref{fdsrapu} we show the distributions of $\Delta y^{+-}$ for
opposite-charge pairs of each of the three combinations of unlike particles,
as well as the sum of the $\Delta y^{++}$ and $\Delta y^{--}$
distributions for the corresponding same-charge pairs.
A significant negative asymmetry is observed for $\pi K$ pairs of both opposite-
and same-charge at all $\Delta y$, which may simply be due the combination of
leading kaons and randomly selected pions.
A similar effect in $K$p pairs can be attributed to leading kaons combined with
randomly chosen protons.  In this case the asymmetry is
negative for opposite-charge pairs and positive for same-charge pairs,
as expected given our sign convention.
In the case of $\pi$p pairs there is a small positive asymmetry for both
opposite- and same-charge pairs.
The asymmetries predicted by the simulation are also shown; they are consistent
with the p$K^+$ data, but overestimate the magnitudes of the effects for
p$K^-$, $K^+\pi^-$ and $K^+\pi^+$ pairs slightly.  The latter two differences
are expected in light of the absence of a $\pi K$ correlation in the simulation
of light-flavor events noted above.
A small negative asymmetry is predicted for $\pi$p pairs, which is inconsistent
with the observed positive asymmetry.

\section{Conclusion}

We have presented a preliminary study of rapidity differences between pairs of
identified
charged pions, kaons and protons in light-flavor hadronic $Z^0$ decays.
The SLD Cherenkov Ring Imaging Detector was used to select clean samples of
identified charged hadrons, and the Vertex Detector was used to investigate the
flavor dependence of the results.

We observe excesses of opposite-charge over same-charge pairs for all pair
combinations 
at low values of the absolute rapidity difference, indicating that there is a
high degree of local conservation
of baryon number, strangeness and electric charge in the fragmentation process.
The predictions of the JETSET fragmentation model are found to provide a
qualitative description of the data, although they fail to describe the forms
of some of the correlations in detail.
The range of these short-range correlations was studied as a function of
momentum; the variations found were reproduced by the simulation, verifying
within the context of the JETSET model that the correlations are scale
invariant.

We observe a large excess of high momentum $K^+K^-$ pairs over same-charge
kaon pairs at large values of the absolute rapidity difference, and 
the effect is larger for higher momenta, as expected from leading kaon
production in $s\bar{s}$ events.
Weaker correlations are observed for protons, indicating that events with
a leading baryon in one jet, a leading antibaryon in the other and no
additional baryons do not contribute significantly to baryon production in
$e^+e^-$ annihilations.
Significant long-range correlations are observed between opposite-charge
pairs of all combinations in light-flavor events;
in $c\bar{c}$ events, long-range
$\pi\pi$, $\pi$p and $KK$ correlations are observed, along with a strong $\pi K$
anticorrelation.
The simulation provides a good description of the data in general, but does not
predict the long-range $\pi K$ correlation in light-flavor events.

We have studied distributions of rapidity signed so that positive rapidity
corresponds to the quark (rather than antiquark) direction.  Differences between
signed rapidity distributions for positive and negative hadrons of all three
species are observed, giving further evidence for leading production of charged
pions kaons and protons.
The distribution of the difference between the signed rapidities of
$K^+$ and $K^-$ shows a large asymmetry at large values of the absolute
rapidity difference, a direct indication that the long-range correlated $KK$
pairs are dominated by $s\bar{s}$ events.
A similar but smaller difference for $\pi^+ \pi^-$ pairs indicates roughly equal
production of leading pions in $u\bar{u}$ and $d\bar{d}$ events.
There is a large asymmetry at small rapidity difference for p$\bar{\rm p}$
pairs, a clear indication of ordering of baryons along the event axis.
A similar effect is observed for $K^+K^-$ pairs at low momentum in
light-flavor events.

\section*{Acknowledgements}
We thank the personnel of the SLAC accelerator department and the
technical
staffs of our collaborating institutions for their outstanding efforts
on our behalf.

\vskip .5truecm

\vbox{\footnotesize\renewcommand{\baselinestretch}{1}\noindent
$^*$Work supported by Department of Energy
  contracts:
  DE-FG02-91ER40676 (BU),
  DE-FG03-91ER40618 (UCSB),
  DE-FG03-92ER40689 (UCSC),
  DE-FG03-93ER40788 (CSU),
  DE-FG02-91ER40672 (Colorado),
  DE-FG02-91ER40677 (Illinois),
  DE-AC03-76SF00098 (LBL),
  DE-FG02-92ER40715 (Massachusetts),
  DE-FC02-94ER40818 (MIT),
  DE-FG03-96ER40969 (Oregon),
  DE-AC03-76SF00515 (SLAC),
  DE-FG05-91ER40627 (Tennessee),
  DE-FG02-95ER40896 (Wisconsin),
  DE-FG02-92ER40704 (Yale);
  National Science Foundation grants:
  PHY-91-13428 (UCSC),
  PHY-89-21320 (Columbia),
  PHY-92-04239 (Cincinnati),
  PHY-95-10439 (Rutgers),
  PHY-88-19316 (Vanderbilt),
  PHY-92-03212 (Washington);
  The UK Particle Physics and Astronomy Research Council
  (Brunel, Oxford and RAL);
  The Istituto Nazionale di Fisica Nucleare of Italy
  (Bologna, Ferrara, Frascati, Pisa, Padova, Perugia);
  The Japan-US Cooperative Research Project on High Energy Physics
  (Nagoya, Tohoku);
  The Korea Research Foundation (Soongsil, 1997).}

\section*{$^{**}$List of Authors} 

%
%
%
\begin{center}
\def\iADEL{$^{(1)}$}
\def\iAOMORI{$^{(2)}$}
\def\iBOLO{$^{(3)}$}
\def\iBRI{$^{(4)}$}
\def\iBRUN{$^{(5)}$}
\def\iBU{$^{(6)}$}
\def\iCINC{$^{(7)}$}
\def\iCOLO{$^{(8)}$}
\def\iCOLU{$^{(9)}$}
\def\iCSU{$^{(10)}$}
\def\iFERR{$^{(11)}$}
\def\iFRAS{$^{(12)}$}
\def\iILLI{$^{(13)}$}
\def\iJHU{$^{(14)}$}
\def\iLBL{$^{(15)}$}
\def\iLTU{$^{(16)}$}
\def\iMASS{$^{(17)}$}
\def\iMISSI{$^{(18)}$}
\def\iMIT{$^{(19)}$}
\def\iMOSCOW{$^{(20)}$}
\def\iNAGO{$^{(21)}$}
\def\iOREG{$^{(22)}$}
\def\iOXF{$^{(23)}$}
\def\iPADO{$^{(24)}$}
\def\iPERU{$^{(25)}$}
\def\iPISA{$^{(26)}$}
\def\iRAL{$^{(27)}$}
\def\iRUTG{$^{(28)}$}
\def\iSLAC{$^{(29)}$}
\def\iSOGA{$^{(30)}$}
\def\iSOONG{$^{(31)}$}
\def\iTENN{$^{(32)}$}
\def\iTOHO{$^{(33)}$}
\def\iUCSB{$^{(34)}$}
\def\iUCSC{$^{(35)}$}
\def\iUVIC{$^{(36)}$}
\def\iVAND{$^{(37)}$}
\def\iWASH{$^{(38)}$}
\def\iWISC{$^{(39)}$}
\def\iYALE{$^{(40)}$}

  \baselineskip=.75\baselineskip  
\mbox{Kenji  Abe\unskip,\iNAGO}
\mbox{Koya Abe\unskip,\iTOHO}
\mbox{T. Abe\unskip,\iSLAC}
\mbox{I.Adam\unskip,\iSLAC}
\mbox{T.  Akagi\unskip,\iSLAC}
\mbox{N. J. Allen\unskip,\iBRUN}
\mbox{W.W. Ash\unskip,\iSLAC}
\mbox{D. Aston\unskip,\iSLAC}
\mbox{K.G. Baird\unskip,\iMASS}
\mbox{C. Baltay\unskip,\iYALE}
\mbox{H.R. Band\unskip,\iWISC}
\mbox{M.B. Barakat\unskip,\iLTU}
\mbox{O. Bardon\unskip,\iMIT}
\mbox{T.L. Barklow\unskip,\iSLAC}
\mbox{G. L. Bashindzhagyan\unskip,\iMOSCOW}
\mbox{J.M. Bauer\unskip,\iMISSI}
\mbox{G. Bellodi\unskip,\iOXF}
\mbox{R. Ben-David\unskip,\iYALE}
\mbox{A.C. Benvenuti\unskip,\iBOLO}
\mbox{G.M. Bilei\unskip,\iPERU}
\mbox{D. Bisello\unskip,\iPADO}
\mbox{G. Blaylock\unskip,\iMASS}
\mbox{J.R. Bogart\unskip,\iSLAC}
\mbox{G.R. Bower\unskip,\iSLAC}
\mbox{J. E. Brau\unskip,\iOREG}
\mbox{M. Breidenbach\unskip,\iSLAC}
\mbox{W.M. Bugg\unskip,\iTENN}
\mbox{D. Burke\unskip,\iSLAC}
\mbox{T.H. Burnett\unskip,\iWASH}
\mbox{P.N. Burrows\unskip,\iOXF}
\mbox{A. Calcaterra\unskip,\iFRAS}
\mbox{D. Calloway\unskip,\iSLAC}
\mbox{B. Camanzi\unskip,\iFERR}
\mbox{M. Carpinelli\unskip,\iPISA}
\mbox{R. Cassell\unskip,\iSLAC}
\mbox{R. Castaldi\unskip,\iPISA}
\mbox{A. Castro\unskip,\iPADO}
\mbox{M. Cavalli-Sforza\unskip,\iUCSC}
\mbox{A. Chou\unskip,\iSLAC}
\mbox{E. Church\unskip,\iWASH}
\mbox{H.O. Cohn\unskip,\iTENN}
\mbox{J.A. Coller\unskip,\iBU}
\mbox{M.R. Convery\unskip,\iSLAC}
\mbox{V. Cook\unskip,\iWASH}
\mbox{R. Cotton\unskip,\iBRUN}
\mbox{R.F. Cowan\unskip,\iMIT}
\mbox{D.G. Coyne\unskip,\iUCSC}
\mbox{G. Crawford\unskip,\iSLAC}
\mbox{C.J.S. Damerell\unskip,\iRAL}
\mbox{M. N. Danielson\unskip,\iCOLO}
\mbox{M. Daoudi\unskip,\iSLAC}
\mbox{N. de Groot\unskip,\iBRI}
\mbox{R. Dell'Orso\unskip,\iPERU}
\mbox{P.J. Dervan\unskip,\iBRUN}
\mbox{R. de Sangro\unskip,\iFRAS}
\mbox{M. Dima\unskip,\iCSU}
\mbox{A. D'Oliveira\unskip,\iCINC}
\mbox{D.N. Dong\unskip,\iMIT}
\mbox{M. Doser\unskip,\iSLAC}
\mbox{R. Dubois\unskip,\iSLAC}
\mbox{B.I. Eisenstein\unskip,\iILLI}
\mbox{V. Eschenburg\unskip,\iMISSI}
\mbox{E. Etzion\unskip,\iWISC}
\mbox{S. Fahey\unskip,\iCOLO}
\mbox{D. Falciai\unskip,\iFRAS}
\mbox{C. Fan\unskip,\iCOLO}
\mbox{J.P. Fernandez\unskip,\iUCSC}
\mbox{M.J. Fero\unskip,\iMIT}
\mbox{K.Flood\unskip,\iMASS}
\mbox{R. Frey\unskip,\iOREG}
\mbox{J. Gifford\unskip,\iUVIC}
\mbox{T. Gillman\unskip,\iRAL}
\mbox{G. Gladding\unskip,\iILLI}
\mbox{S. Gonzalez\unskip,\iMIT}
\mbox{E. R. Goodman\unskip,\iCOLO}
\mbox{E.L. Hart\unskip,\iTENN}
\mbox{J.L. Harton\unskip,\iCSU}
\mbox{A. Hasan\unskip,\iBRUN}
\mbox{K. Hasuko\unskip,\iTOHO}
\mbox{S. J. Hedges\unskip,\iBU}
\mbox{S.S. Hertzbach\unskip,\iMASS}
\mbox{M.D. Hildreth\unskip,\iSLAC}
\mbox{J. Huber\unskip,\iOREG}
\mbox{M.E. Huffer\unskip,\iSLAC}
\mbox{E.W. Hughes\unskip,\iSLAC}
\mbox{X.Huynh\unskip,\iSLAC}
\mbox{H. Hwang\unskip,\iOREG}
\mbox{M. Iwasaki\unskip,\iOREG}
\mbox{D. J. Jackson\unskip,\iRAL}
\mbox{P. Jacques\unskip,\iRUTG}
\mbox{J.A. Jaros\unskip,\iSLAC}
\mbox{Z.Y. Jiang\unskip,\iSLAC}
\mbox{A.S. Johnson\unskip,\iSLAC}
\mbox{J.R. Johnson\unskip,\iWISC}
\mbox{R.A. Johnson\unskip,\iCINC}
\mbox{T. Junk\unskip,\iSLAC}
\mbox{R. Kajikawa\unskip,\iNAGO}
\mbox{M. Kalelkar\unskip,\iRUTG}
\mbox{Y. Kamyshkov\unskip,\iTENN}
\mbox{H.J. Kang\unskip,\iRUTG}
\mbox{I. Karliner\unskip,\iILLI}
\mbox{H. Kawahara\unskip,\iSLAC}
\mbox{Y. D. Kim\unskip,\iSOGA}
\mbox{M.E. King\unskip,\iSLAC}
\mbox{R. King\unskip,\iSLAC}
\mbox{R.R. Kofler\unskip,\iMASS}
\mbox{N.M. Krishna\unskip,\iCOLO}
\mbox{R.S. Kroeger\unskip,\iMISSI}
\mbox{M. Langston\unskip,\iOREG}
\mbox{A. Lath\unskip,\iMIT}
\mbox{D.W.G. Leith\unskip,\iSLAC}
\mbox{V. Lia\unskip,\iMIT}
\mbox{C.Lin\unskip,\iMASS}
\mbox{M.X. Liu\unskip,\iYALE}
\mbox{X. Liu\unskip,\iUCSC}
\mbox{M. Loreti\unskip,\iPADO}
\mbox{A. Lu\unskip,\iUCSB}
\mbox{H.L. Lynch\unskip,\iSLAC}
\mbox{J. Ma\unskip,\iWASH}
\mbox{G. Mancinelli\unskip,\iRUTG}
\mbox{S. Manly\unskip,\iYALE}
\mbox{G. Mantovani\unskip,\iPERU}
\mbox{T.W. Markiewicz\unskip,\iSLAC}
\mbox{T. Maruyama\unskip,\iSLAC}
\mbox{H. Masuda\unskip,\iSLAC}
\mbox{E. Mazzucato\unskip,\iFERR}
\mbox{A.K. McKemey\unskip,\iBRUN}
\mbox{B.T. Meadows\unskip,\iCINC}
\mbox{G. Menegatti\unskip,\iFERR}
\mbox{R. Messner\unskip,\iSLAC}
\mbox{P.M. Mockett\unskip,\iWASH}
\mbox{K.C. Moffeit\unskip,\iSLAC}
\mbox{T.B. Moore\unskip,\iYALE}
\mbox{M.Morii\unskip,\iSLAC}
\mbox{D. Muller\unskip,\iSLAC}
\mbox{V.Murzin\unskip,\iMOSCOW}
\mbox{T. Nagamine\unskip,\iTOHO}
\mbox{S. Narita\unskip,\iTOHO}
\mbox{U. Nauenberg\unskip,\iCOLO}
\mbox{H. Neal\unskip,\iSLAC}
\mbox{M. Nussbaum\unskip,\iCINC}
\mbox{N.Oishi\unskip,\iNAGO}
\mbox{D. Onoprienko\unskip,\iTENN}
\mbox{L.S. Osborne\unskip,\iMIT}
\mbox{R.S. Panvini\unskip,\iVAND}
\mbox{C. H. Park\unskip,\iSOONG}
\mbox{T.J. Pavel\unskip,\iSLAC}
\mbox{I. Peruzzi\unskip,\iFRAS}
\mbox{M. Piccolo\unskip,\iFRAS}
\mbox{L. Piemontese\unskip,\iFERR}
\mbox{K.T. Pitts\unskip,\iOREG}
\mbox{R.J. Plano\unskip,\iRUTG}
\mbox{R. Prepost\unskip,\iWISC}
\mbox{C.Y. Prescott\unskip,\iSLAC}
\mbox{G.D. Punkar\unskip,\iSLAC}
\mbox{J. Quigley\unskip,\iMIT}
\mbox{B.N. Ratcliff\unskip,\iSLAC}
\mbox{T.W. Reeves\unskip,\iVAND}
\mbox{J. Reidy\unskip,\iMISSI}
\mbox{P.L. Reinertsen\unskip,\iUCSC}
\mbox{P.E. Rensing\unskip,\iSLAC}
\mbox{L.S. Rochester\unskip,\iSLAC}
\mbox{P.C. Rowson\unskip,\iCOLU}
\mbox{J.J. Russell\unskip,\iSLAC}
\mbox{O.H. Saxton\unskip,\iSLAC}
\mbox{T. Schalk\unskip,\iUCSC}
\mbox{R.H. Schindler\unskip,\iSLAC}
\mbox{B.A. Schumm\unskip,\iUCSC}
\mbox{J. Schwiening\unskip,\iSLAC}
\mbox{S. Sen\unskip,\iYALE}
\mbox{V.V. Serbo\unskip,\iSLAC}
\mbox{M.H. Shaevitz\unskip,\iCOLU}
\mbox{J.T. Shank\unskip,\iBU}
\mbox{G. Shapiro\unskip,\iLBL}
\mbox{D.J. Sherden\unskip,\iSLAC}
\mbox{K. D. Shmakov\unskip,\iTENN}
\mbox{C. Simopoulos\unskip,\iSLAC}
\mbox{N.B. Sinev\unskip,\iOREG}
\mbox{S.R. Smith\unskip,\iSLAC}
\mbox{M. B. Smy\unskip,\iCSU}
\mbox{J.A. Snyder\unskip,\iYALE}
\mbox{H. Staengle\unskip,\iCSU}
\mbox{A. Stahl\unskip,\iSLAC}
\mbox{P. Stamer\unskip,\iRUTG}
\mbox{H. Steiner\unskip,\iLBL}
\mbox{R. Steiner\unskip,\iADEL}
\mbox{M.G. Strauss\unskip,\iMASS}
\mbox{D. Su\unskip,\iSLAC}
\mbox{F. Suekane\unskip,\iTOHO}
\mbox{A. Sugiyama\unskip,\iNAGO}
\mbox{S. Suzuki\unskip,\iNAGO}
\mbox{M. Swartz\unskip,\iJHU}
\mbox{A. Szumilo\unskip,\iWASH}
\mbox{T. Takahashi\unskip,\iSLAC}
\mbox{F.E. Taylor\unskip,\iMIT}
\mbox{J. Thom\unskip,\iSLAC}
\mbox{E. Torrence\unskip,\iMIT}
\mbox{N. K. Toumbas\unskip,\iSLAC}
\mbox{T. Usher\unskip,\iSLAC}
\mbox{C. Vannini\unskip,\iPISA}
\mbox{J. Va'vra\unskip,\iSLAC}
\mbox{E. Vella\unskip,\iSLAC}
\mbox{J.P. Venuti\unskip,\iVAND}
\mbox{R. Verdier\unskip,\iMIT}
\mbox{P.G. Verdini\unskip,\iPISA}
\mbox{D. L. Wagner\unskip,\iCOLO}
\mbox{S.R. Wagner\unskip,\iSLAC}
\mbox{A.P. Waite\unskip,\iSLAC}
\mbox{S. Walston\unskip,\iOREG}
\mbox{J.Wang\unskip,\iSLAC}
\mbox{S.J. Watts\unskip,\iBRUN}
\mbox{A.W. Weidemann\unskip,\iTENN}
\mbox{E. R. Weiss\unskip,\iWASH}
\mbox{J.S. Whitaker\unskip,\iBU}
\mbox{S.L. White\unskip,\iTENN}
\mbox{F.J. Wickens\unskip,\iRAL}
\mbox{B. Williams\unskip,\iCOLO}
\mbox{D.C. Williams\unskip,\iMIT}
\mbox{S.H. Williams\unskip,\iSLAC}
\mbox{S. Willocq\unskip,\iMASS}
\mbox{R.J. Wilson\unskip,\iCSU}
\mbox{W.J. Wisniewski\unskip,\iSLAC}
\mbox{J. L. Wittlin\unskip,\iMASS}
\mbox{M. Woods\unskip,\iSLAC}
\mbox{G.B. Word\unskip,\iVAND}
\mbox{T.R. Wright\unskip,\iWISC}
\mbox{J. Wyss\unskip,\iPADO}
\mbox{R.K. Yamamoto\unskip,\iMIT}
\mbox{J.M. Yamartino\unskip,\iMIT}
\mbox{X. Yang\unskip,\iOREG}
\mbox{J. Yashima\unskip,\iTOHO}
\mbox{S.J. Yellin\unskip,\iUCSB}
\mbox{C.C. Young\unskip,\iSLAC}
\mbox{H. Yuta\unskip,\iAOMORI}
\mbox{G. Zapalac\unskip,\iWISC}
\mbox{R.W. Zdarko\unskip,\iSLAC}
\mbox{J. Zhou\unskip.\iOREG}

\it
  \vskip \baselineskip                   
  \vskip \baselineskip        
  \baselineskip=.75\baselineskip   
\iADEL
  Adelphi University, Garden City, New York 11530, \break
\iAOMORI
  Aomori University, Aomori , 030 Japan, \break
\iBOLO
  INFN Sezione di Bologna, I-40126, Bologna Italy, \break
\iBRI
  University of Bristol, Bristol, U.K., \break
\iBRUN
  Brunel University, Uxbridge, Middlesex, UB8 3PH United Kingdom, \break
\iBU
  Boston University, Boston, Massachusetts 02215, \break
\iCINC
  University of Cincinnati, Cincinnati, Ohio 45221, \break
\iCOLO
  University of Colorado, Boulder, Colorado 80309, \break
\iCOLU
  Columbia University, New York, New York 10533, \break
\iCSU
  Colorado State University, Ft. Collins, Colorado 80523, \break
\iFERR
  INFN Sezione di Ferrara and Universita di Ferrara, I-44100 Ferrara, Italy, \break
\iFRAS
  INFN Lab. Nazionali di Frascati, I-00044 Frascati, Italy, \break
\iILLI
  University of Illinois, Urbana, Illinois 61801, \break
\iJHU
  Johns Hopkins University, Baltimore, MD 21218-2686, \break
\iLBL
  Lawrence Berkeley Laboratory, University of California, Berkeley, California 94720, \break
\iLTU
  Louisiana Technical University - Ruston,LA 71272, \break
\iMASS
  University of Massachusetts, Amherst, Massachusetts 01003, \break
\iMISSI
  University of Mississippi, University, Mississippi 38677, \break
\iMIT
  Massachusetts Institute of Technology, Cambridge, Massachusetts 02139, \break
\iMOSCOW
  Institute of Nuclear Physics, Moscow State University, 119899, Moscow Russia, \break
\iNAGO
  Nagoya University, Chikusa-ku, Nagoya 464 Japan, \break
\iOREG
  University of Oregon, Eugene, Oregon 97403, \break
\iOXF
  Oxford University, Oxford, OX1 3RH, United Kingdom, \break
\iPADO
  INFN Sezione di Padova and Universita di Padova I-35100, Padova, Italy, \break
\iPERU
  INFN Sezione di Perugia and Universita di Perugia, I-06100 Perugia, Italy, \break
\iPISA
  INFN Sezione di Pisa and Universita di Pisa, I-56010 Pisa, Italy, \break
\iRAL
  Rutherford Appleton Laboratory, Chilton, Didcot, Oxon OX11 0QX United Kingdom, \break
\iRUTG
  Rutgers University, Piscataway, New Jersey 08855, \break
\iSLAC
  Stanford Linear Accelerator Center, Stanford University, Stanford, California 94309, \break
\iSOGA
  Sogang University, Seoul, Korea, \break
\iSOONG
  Soongsil University, Seoul, Korea 156-743, \break
\iTENN
  University of Tennessee, Knoxville, Tennessee 37996, \break
\iTOHO
  Tohoku University, Sendai 980, Japan, \break
\iUCSB
  University of California at Santa Barbara, Santa Barbara, California 93106, \break
\iUCSC
  University of California at Santa Cruz, Santa Cruz, California 95064, \break
\iUVIC
  University of Victoria, Victoria, B.C., Canada, V8W 3P6, \break
\iVAND
  Vanderbilt University, Nashville,Tennessee 37235, \break
\iWASH
  University of Washington, Seattle, Washington 98105, \break
\iWISC
  University of Wisconsin, Madison,Wisconsin 53706, \break
\iYALE
  Yale University, New Haven, Connecticut 06511. \break

\rm
%

\end{center}



\begin{thebibliography}{99}
 
\renewcommand{\baselinestretch}{1.0}  
  
\bibitem{srcor}
TPC Collab., H.~Aihara et al., Phys. Rev. Lett. {\bf 57} (1986) 3140;  \\
TASSO Collab., R.~Brandelik et al., Phys. Lett. {\bf B139} (1984) 126;  \\
ALEPH Collab., D.~Buskulic et al., Z. Phys. {\bf C64}~(1994)~361;  \\
DELPHI Collab., P.~Abreu et al., Phys. Lett. {\bf B416} (1998) 247;  \\
OPAL Collab., P.D.~Acton et al., Phys. Lett. {\bf B305} (1993) 415.
 
\bibitem{lrcor}
TPC Collab., H.~Aihara et al., Phys. Rev. Lett. {\bf 53} (1984) 2199;  \\
TASSO Collab., R.~Brandelik et al., Phys. Lett. {\bf B100} (1981) 357;  \\
A. Breakstone et al., Z. Phys. {\bf C25} (1984) 21.
 
\bibitem{sld} SLD Design Report, SLAC-Report 273 (1984).

\bibitem{sldalphas}
SLD Collaboration: K.~Abe et  al., Phys.\ Rev.\ {\bf D51} (1995) 962.

\bibitem{cdc} 
M.~D.~Hildreth et al.,  Nucl.\ Instr.\ Meth. {\bf A367} (1995) 111.

\bibitem {vxd} 
C.J.S. Damerell et. al., Nucl.\ Instr.\ Meth. {\bf A288} (1990) 236. \\
C.J.S. Damerell et. al., Nucl.\ Instr.\ Meth. {\bf A400} (1997) 287.

\bibitem{crid}
K. Abe et al., Nucl. Inst. Meth. {\bf A343} (1994) 74.

\bibitem{bfp}
SLD Collab., K. Abe et al., Phys. Rev. {\bf D59} 52001.

\bibitem {thrust}
S. Brandt et al., Phys. Lett. {\bf 12}~(1964)~57.\\
E. Farhi, Phys. Rev. Lett. {\bf 39}~(1977)~1587.

\bibitem {lac} 
D. Axen et al., Nucl. Inst. Meth. {\bf A328}~(1993)~472.

\bibitem{homer}
SLD Collab., K.~Abe et al., Phys. Rev. {\bf D53} (1996) 1023.

\bibitem{jetset} 
T.~Sj\"ostrand, Comput.\ Phys.\ Commun.\ {\bf 82} (1994) 74.
 
\bibitem{tune}
P. N. Burrows, Z. Phys. {\bf C41} (1988) 375.\\
OPAL Collaboration, M.Z. Akrawy et al., Z. Phys. {\bf C47} (1990) 505.
 
\bibitem{sldsim}
SLD Collaboration, K. Abe et al., Phys. Rev. Lett. {\bf 79} (1997) 590.

\bibitem{geant}
R. Brun et al., Report No. CERN-DD/EE/84-1 (1989). 
 
\bibitem{emud}
SLD Collaboration, K. Abe et al., Phys. Rev. Lett. {\bf 74} (1995) 2895.

\bibitem{llik}
K. Abe et al., Nucl. Inst. and Meth. {\bf A371} (1996) 195.

\bibitem{lpprl} SLD Collab., K. Abe et al.,
Phys. Rev. Lett. {\bf 78} (1997) 3442.

\end{thebibliography}
\end{document}